\documentclass[prx,aps,showpacs,twocolumn,amsmath,amssymb,floatfix,superscriptaddress]{revtex4-1}

\usepackage{hyperref}
\usepackage{graphicx}
\usepackage{bm}
\usepackage{color}
\usepackage[applemac]{inputenc}
\usepackage{amssymb}

\newcommand{\beq}{\begin{equation}}
\newcommand{\eeq}{\end{equation}}
\newcommand{\beqa}{\begin{eqnarray}}
\newcommand{\eeqa}{\end{eqnarray}}
\newcommand{\sect}[1]{\emph{#1.---}\ignorespaces} 

\newcommand{\G}{\mathcal{G}} 
\newcommand{\T}{\mathcal{T}} 
\newcommand{\dr}{\delta\bm{r}}
\renewcommand{\r}{{\bm{r}}}
\renewcommand{\k}{{\bm{k}}}
\newcommand{\g}{{\bm{g}}}
\newcommand{\bra}[1]{\langle{#1}|}
\newcommand{\ket}[1]{|{#1}\rangle}
\newcommand{\braket}[2]{\langle{#1}|{#2}\rangle}
\newcommand{\intk}{\int \frac{d^2 k}{(2\pi)^2}}

\newcommand{\edit}[1] {\textcolor{black}{#1}}

\begin{document}

\title{Flat bands in magic-angle vibrating plates}

\author{Mar\'ia Rosendo}
\affiliation{Department of Physics, Universidad Carlos III de Madrid, ES-28916 Legan\`es, Madrid, Spain}
\author{Fernando Pe\~naranda}
\affiliation{Instituto de Ciencia de Materiales de Madrid, Consejo Superior de Investigaciones Cient\'{i}ficas (ICMM-CSIC), Madrid, Spain}
\author{Johan Christensen}
\affiliation{Department of Physics, Universidad Carlos III de Madrid, ES-28916 Legan\`es, Madrid, Spain}
\author{Pablo San-Jose}
\affiliation{Instituto de Ciencia de Materiales de Madrid, Consejo Superior de Investigaciones Cient\'{i}ficas (ICMM-CSIC), Madrid, Spain}
 
\date{\today}
 
\begin{abstract}
Twisted bilayer graphene develop quasi-flat bands at specific ``magic'' interlayer rotation angles through an unconventional mechanism connected to carrier chirality. Quasi-flat bands are responsible for a wealth of exotic, correlated-electron phases in the system. In this work we propose a mechanical analogue of twisted bilayer graphene made of two vibrating plates, patterned with a honeycomb mesh of masses, and coupled across a continuum elastic medium. We show that flexural waves in the device exhibit vanishing group velocity and quasi-flat bands at magic angles, in close correspondence with electrons in graphene models. The strong similarities of spectral structure and spatial eigenmodes in the two systems demonstrate the chiral nature of the mechanical flat bands. We derive analytical expressions that quantitatively connect the mechanical and electronic models, which allow us to predict the parameters required for an experimental realization of our proposal. 
\end{abstract}

\maketitle

Classical analogues of quantum electronic systems in acoustic and mechanical settings offer a new and exciting perspective on non-trivial electronic phenomena, such as topological insulating phases, topologically protected edge states, Weyl and Dirac semimetallic phases or Majorana bound states \cite{ZHONG20113533,Torrent:PRB13,he2016acoustic,zangeneh2019topological,chen2019mechanical,gao2019majorana,chaunsali2017demonstrating,miniaci2018experimental,chen2019topological}. An important appeal of these classical analogues is their easy fabrication and tuneability, typically much simpler than for their electronic counterparts. They often reveal new and unexpected effects in a classical context and deep connections between very different physical systems \cite{zhang2018topologicalsound}.

\begin{figure}
   \centering
   \includegraphics[width=\columnwidth]{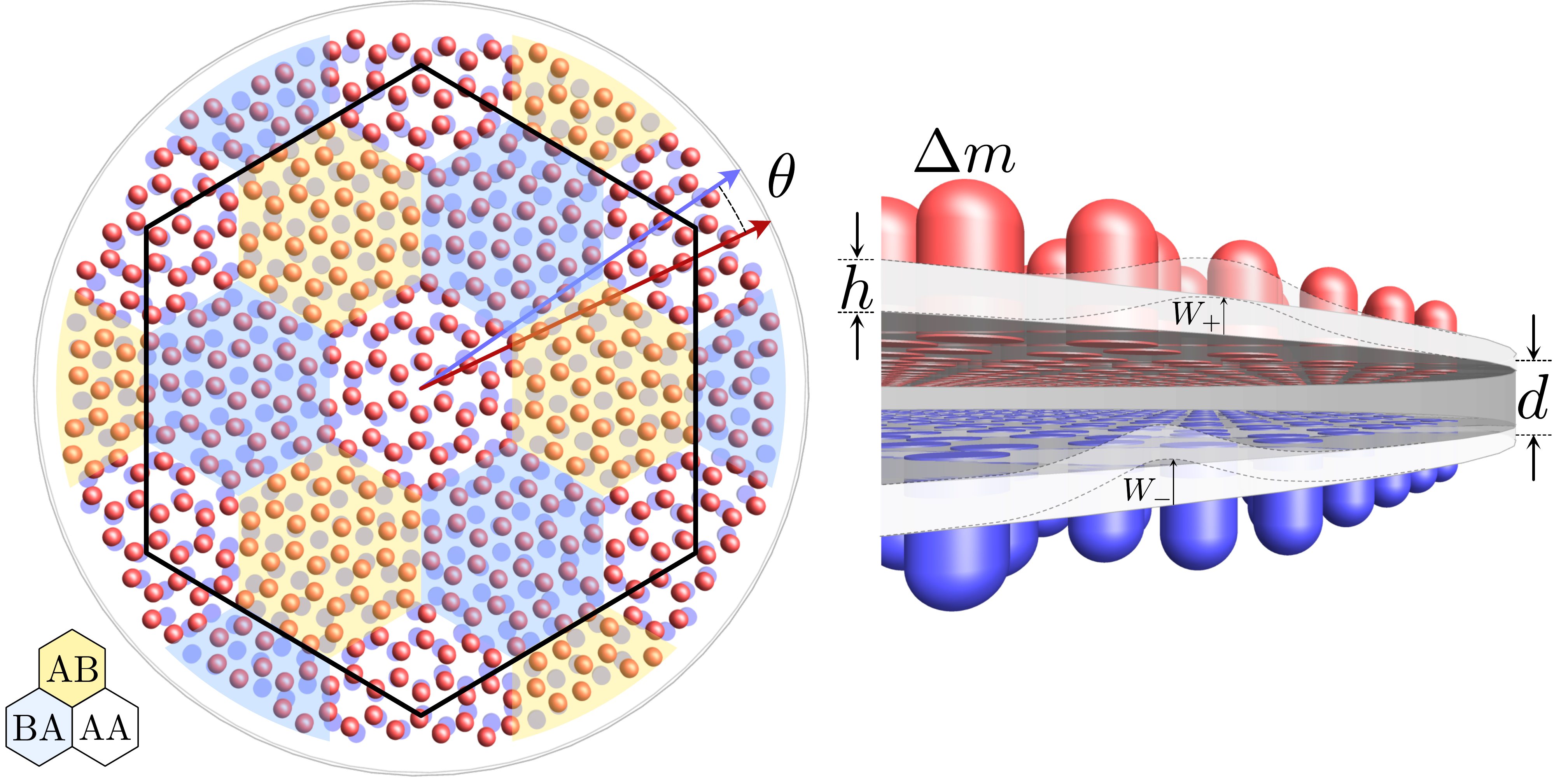}
   \caption{Mechanical analogue of twisted bilayer graphene: \edit{two vibrating plates of thickness $h$, density $\rho$, Young modulus $E$ and Poisson ratio $\nu$, patterned with a honeycomb lattice of $\Delta m$ masses (blue and red) and lattice constant $a$, and coupled across an elastic medium of thickness $d$ and Young modulus $E_d$. Upon changing the relative plate rotation angle $\theta$, a moir\'e pattern of alternating AA/AB/BA stacking alignments emerge, which modifies the group velocity of out-of-plane (flexural) modes ($W_\pm$). At specific (``magic'') angles $\theta_i$, the group velocity vanishes, and quasi-flat flexural bands emerge. A possible realistic implementation could use LiNbO${}_3$ plates and a rubber spacer, with $h\approx1\mu$m, $\rho \approx 4640$kg/m${}^3$, $E\approx 170$GPa, $\nu\approx0.25$, $\Delta m \approx 5.8$ng, $a\approx 24\mu$m, $d\approx5\mu$m and $E_d\approx 10$ MPa. This yields dimensionless parameters $\gamma\approx 2.5$, $\kappa\approx30$, and $\Omega_D\approx6.3$ (around 20 MHz), see Eq. \eqref{dimpar}, and a first magic angle at $\theta_1\approx1.6^\circ$.}}
   \label{fig:1}
\end{figure}

A remarkable electronic effect that has to date received little attention in the acoustic and mechanical context is flat-band formation in twisted bilayer graphene (TBG). TBG is composed of two graphene monolayers placed in direct contact with each other after rotating one of them by a certain angle $\theta$ \cite{Santos:PRL07,Suarez-Morell:PRB10,Bistritzer:PNAS11,Santos:PRB12}. Each monolayer on its own possesses a massless Dirac spectrum with a certain group velocity $v_0$ around Dirac wavevectors $\pm\bf{K}$ \cite{Neto:RMP09,Amorim:PR16}. The crystalline moir\'e pattern produced by the interlayer rotation, Fig. \ref{fig:1}a, was shown \cite{Santos:PRL07,Luican:PRL11} to produce a $\theta$-dependent suppression of the velocity $v(\theta)$, even reaching $v(\theta_i)=0$ at a series of so-called magic angles $\theta_{i=1,2,\dots}$ \cite{Suarez-Morell:PRB10,Bistritzer:PNAS11,Trambly-de-Laissardiere:PRB12,Moon:PRB12}. At these twist angles, the TBG Dirac cones degenerate into quasi-flat bands at the half-filling Fermi energy. The mechanism behind flat-band formation in the system is highly unconventional, and is not the result of exponential wavefunction localization (although algebraic localization at AA moir\'e region does takes place \cite{Laissardiere:NL10}), but of carrier chirality and effective non-Abelian gauge fields produced by the modulation of the interlayer coupling \cite{San-Jose:PRL12,Tarnopolsky:PRL19}. The development of chirality-driven quasi-flat bands produces a rich phase diagram of correlated electronic phases triggered by many-body instabilities\cite{Vafek:PRB10,Isobe:PRX18,Kennes:PRB18,Gonzalez:PRL19,Sboychakov:PRB19,Andrei:20}, which include Mott-insulating phases\cite{Cao:N18b,Lu:N19}, non-conventional superconductivity (possibly related to that of cuprates) \cite{Cao:N18,Liu:PRL18,Yankowitz:S19,Lu:N19}, strange-metal behaviour \cite{Cao:PRL20,Lyu:20} and two-dimensional magnetism\cite{Gonzalez-Arraga:PRL17,Thomson:PRB18,Cao:N18,Lu:N19,Sharpe:S19,Zondiner:N20}. These correlated phases are experimentally found to emerge at the first magic angle, and are thus generally understood as a non-trivial consequence of quasi-flat band formation.

In this work we propose a mechanical analogue of TBG consisting of two elastic plates, supporting flexural (out-of-plane) vibrations. The plates are homogeneously coupled across a thin elastic medium, and a honeycomb pattern of point-like masses is attached to each, see Fig. \ref{fig:1}. We demonstrate a strong modulation of the flexural wave group velocity with the inter-plate rotation angle $\theta$, and the emergence of quasi-flat flexural-mode bands at magic angles, in close correspondence with the electronic counterpart. We showcase these effects by numerically solving the multiple-scattering problem of flexural modes on the attached masses as a function of $\theta$. The freezing of flexural vibrations into quasi-flat bands happens at specific magic angles that in turn depend on mechanical parameters. We also derive approximate analytical expressions that connect the mechanical description of our system to the canonical electronic models used for TGB, establishing a precise connection between the two. The mapping allows us to directly compare the different spatial structure of eigenstates in equivalent mechanical and electronic systems. \edit{A realistic experimental implementation of our proposal is possible, with an example of fabrication parameters summarized in Fig. \ref{fig:1}.}

\begin{figure}
   \centering
   \includegraphics[width=\columnwidth]{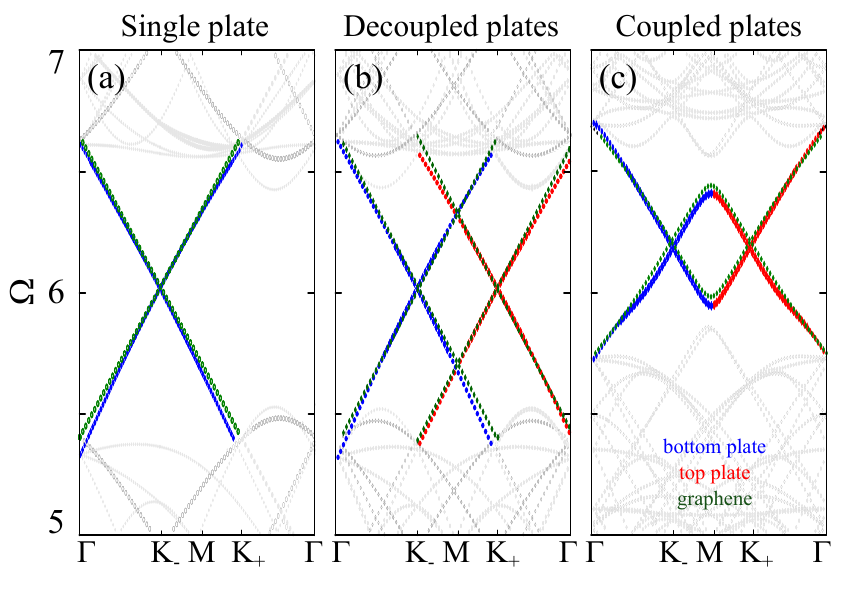}
   \caption{Dirac cones in the normalized bandstructure $\Omega(\k)$ of a single patterned plate (a), two decoupled ($\kappa=0$) but rotated ($m=5, \theta\approx6^\circ$) plates (b), and two coupled ($\kappa=20$) and rotated plates (c). Red/blue denote eigenvalues mostly concentrated on the top/bottom layers, whose respective Dirac points are located at $K_\pm$. The anticrossing at the $M$ point is a van Hove singularity. In green, the normalized bandstructure of the equivalent graphene counterparts.}
   \label{fig:2}
\end{figure}

\begin{figure}
   \centering
   \includegraphics[width=\columnwidth]{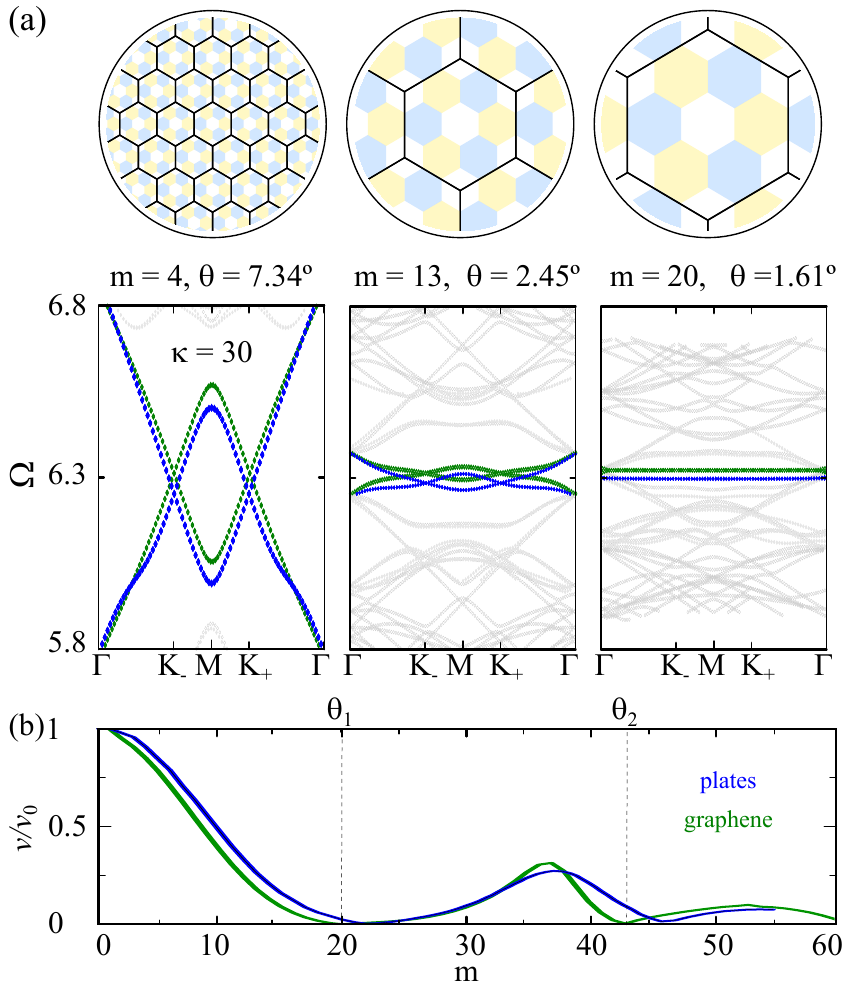}
   \caption{(a) Flat band formation as the angle is decreased towards the first magic angle for a double plate system with $\gamma=2.5$ and $\kappa=30$ (blue), and for the equivalent twisted bilayer graphene system (green). (b) Evolution of the group velocity $v$ at the Dirac point with twist angle $\theta$, normalized to the Dirac velocity $v_0$ of decoupled layers. The angles with vanishing velocity define so-called magic angles $\theta_n$. For the chosen parameters, $\theta_1=1.61^\circ$.}
   \label{fig:3}
\end{figure}

\sect{Structured double plates}
Consider flexural waves with amplitudes $W_l$ in two thin plates $l=\pm$ of uniform mass density $\rho$, thickness $h$, Young modulus $E$, Poisson ratio $\nu$ and bending stiffness $D=Eh^3/[12(1-\nu^2)]$. The vibrations of the two layers are elastically coupled locally by a linear intermediate medium of thickness $d$ and Young modulus $E_d$. We structure each plate with a honeycomb lattice of point masses represented by a mass density perturbation $\delta\rho_l(\r)$ on plate $l$, see sketch in Fig. \ref{fig:1}. The equation of motion governing the flexural waves in the system can be approximated by two coupled Germain-Lagrange equations. In the frequency $\omega$ domain,
\beqa
\label{eom}
\sum_{l'}\left[\left(h\rho\omega^2-D\nabla^4 - \frac{E_d}{d}\right)\tau_0^{ll'} + \frac{E_d}{d}\tau_x^{ll'}\right]W_{l'}(\r,\omega) \nonumber\\
= -h\omega^2\delta\rho_l(\r) W_l(\r,\omega)
\eeqa
Here, the Pauli matrices $\tau_x$ and $\tau_0$ act on the ``layer'' (plate) index $l$. The rotation angle between layers enters into the mass density perturbation $\delta\rho_l(\r)$, which we write as
\beq
\delta\rho_l(\r)=\sum_{\alpha=A,B}\sum_{\r^l_\alpha} \frac{\Delta m}{A_c h}\delta(\r-\r^l_{\alpha}),
\eeq
where $\r^l_{A,B}=n_1\bm{a}^l_1+n_2\bm{a}^l_2\mp (\bm{a}^l_1+\bm{a}^l_2)/6$ for integer $n_{1,2}$ denotes the positions of the point masses $\Delta m$ in layer $l=\pm$. The point masses form a honeycomb lattice with Bravais vectors $\bm{a}^l_{1,2}=a\,\mathcal{U}(l\,\theta/2)\left[\pm \cos(\pi/3),\sin(\pi/3)\right]$ on each layer $l$, with $\mathcal{U}(\theta)$ the relative rotation between layers and $a$ the honeycomb lattice period. $A_c=\sqrt{3}a^2/2$ stands for the area of the honeycomb unit cell. For our numerics, we restrict $\theta$ to commensurate rotations $\theta = \arccos\left[(3m^2 + 3m + 1/2)/(3m^2 + 3m + 1)\right]$ for some integer $m$. Under this constraint the moir\'e pattern resulting from overlapping the two plates is exactly periodic, with a period $L_m=a/[2\sin(\theta/2)]$. 

To compare to the TBG case, it is useful to recast Eq. \ref{eom} into a dimensionless form
\beqa
\label{eom2}
\sum_{l'}\left[\left(\Omega^2-a^4\nabla^4 - \kappa\right)\tau_0^{ll'} + \kappa\tau_x^{ll'}\right]w_{l'}(\r,\omega) \nonumber\\
= -\gamma\Omega^2\sum_{\alpha=A,B}\sum_{\r^l_\alpha} \delta(\r-\r^l_{\alpha}) w_l(\r,\omega)
\eeqa
where we have introduced the dimensionless vibration amplitude $w = \frac{D}{a^4}W$ and dimensionless constants
\begin{eqnarray}
\label{dimpar}
\Omega^2 = \frac{a^4h\rho\omega^2}{D} \hspace{0.5cm}
\kappa = \frac{a^4E_d}{Dd} \hspace{0.5cm}
\gamma = \frac{\Delta m}{\rho h A_c}
\end{eqnarray}

\sect{Results} Solving the eigenvalues $\Omega$ of Eq. \eqref{eom2} in wavevector $\k$ space in the large angle regime (see Appendix \ref{ap:1}) we obtain the wave dispersions shown in Fig. \ref{fig:2}. They are presented along a cut $\Gamma K_- M K_+ \Gamma$ of the moir\'e Brillouin zone for three distinct cases. The $\bm K_l$ Dirac wavevectors of the two layers are located at $\bm{K}_\pm=\mathcal{U}(\pm\theta/2)(4\pi/3a,0)$. In (a) we show the solution for a single plate $l=-$. The dispersion clearly shows the emergence of a Dirac cone at $\k=\bm{K}_-$ (blue lines) around $\Omega=\Omega_D\approx 6$ when the mass lattice is added to the plate \cite{ZHONG20113533,Torrent:PRB13}. Panel (b) shows the spectrum for two decoupled plates ($\kappa=0$) with a relative $\theta$ rotation. The Dirac cone of the second $l=+$ plate (red lines) appears at momentum $\k=\bm{K}_+$ and crosses the one from the $l=-$ plate at the $M$ point. Finally, panel (c) shows the spectrum for the two plates coupled by a finite plate coupling $\kappa$. An anticrossing between the two Dirac cone emerges, producing a van-Hove singularity in the density of states of the system. This is exactly the phenomenology predicted \cite{Santos:PRL07} and observed \cite{Luican:PRL11} for twisted bilayer graphene at not-so-small angles, $\theta\gtrsim 3^\circ$. Note, however, that the formulation of the system model is very different from that of TBG. In contrast to the wave equation Eq. \eqref{eom2}, TBG is usually described using the TBG continuum Hamiltonian \cite{Santos:PRB12}, which can be succinctly written as \cite{San-Jose:PRL12}
\begin{eqnarray}
\label{TBGmodel}
H(\k) &=& \left(\begin{array}{cccc}
t_0\Omega_D & \Pi^\dagger_+ & V_\mathrm{AA}(\r)^* & V_\mathrm{BA}(\r)^*\\
\Pi_+ & t_0\Omega_D & V_\mathrm{AB}(\r)^* & V_\mathrm{AA}(\r)^*\\
V_\mathrm{AA}(\r) & V_\mathrm{AB}(\r) & t_0\Omega_D & \Pi^\dagger_-\\
V_\mathrm{BA}(\r) & V_\mathrm{AA}(\r) & \Pi_- & t_0\Omega_D
\end{array}\right)\hspace{.4cm}\\
\Pi_\pm &=& (k_x+ik_y\mp i\Delta K/2)v_0\nonumber\\
\Delta K &=& |\bm{K}_+-\bm{K}_-|=\frac{4\pi}{3a}2\sin\frac{\theta}{2}\nonumber = \frac{4\pi}{3L_m}\\
V_\alpha &=& \frac{t_\perp}{3}\left(1+e^{i \bm{G}_1 (\r-\r_\alpha)}+e^{i \bm{G}_2 (\r-\r_\alpha)}\right)\nonumber\\
\r_\mathrm{AA}(\r) &=& 0,\hspace{0.4cm} \r_\mathrm{AB}=-\r_\mathrm{BA}=\left(\frac{L_m}{\sqrt{3}},0\right).\nonumber
\end{eqnarray}
The TBG bandstructure $\epsilon(\k)\equiv  t_0\Omega(\k)$ and the corresponding eigenstates $\psi(\k)$ is obtain from the eigenvalue equation $H(\k)\psi(\k) = \epsilon(\k)\psi(\k)$. The model exhibits the explicit $4\times 4$ pseudospin-layer structure of a bilayer Dirac system, unlike the plate equation Eq. \eqref{eom2}. The parameters specific to this model are the  energy scale $t_0\approx 2.7$eV (intralayer hopping amplitude, or one third of the monolayer bandwidth), the twist angle $\theta$ or period $L_m$ \big[which enters through the moir\'e momenta $\bm{G}_{1,2}=\frac{2\pi}{L_m}\left(\pm\frac{1}{\sqrt{3}},1\right)$\big], the Dirac velocity $v_0$ of the decoupled layers, the Fermi energy at half-filling $t_0\Omega_D$ and the interlayer hopping $t_\perp$ (whose moir\'e-induced modulation in the plane is captured by the $V_\alpha(\r)$ functions). $t_\perp$ plays a role analogous to $\kappa$ in the coupled plates, although in the latter the coupling is spatially uniform. The above TBG model neglects any particle-hole asymmetries in the decoupled layers around the Dirac point, which would arise in particular from finite next-nearest-neighbor hoppings in-plane. It also assumes negligible layer strains and identical $V_\alpha$ coefficients. Both conditions are sometimes relaxed in more elaborate versions of the model.

We can formally connect the mechanical and electronic models analytically. The mapping is valid to second order in the effective plate coupling $\kappa/|a\bm{K}|^4$. The starting point is a projection of the plane wave basis into a ``tight-binding'' basis of flexural modes spatially localized at the point masses on each layer. By carefully integrating out the remaining plate vibrations between scatterers in Eq. \eqref{eom2}, the continuum TGB model of Eq. \eqref{TBGmodel} emerges for frequencies close to the Dirac point and small couplings $\kappa$. The detailed demonstration is presented in full in Appendix \ref{ap:2}. Here we present only the final result connecting the plate parameters in Eq. \eqref{dimpar} to the equivalent TBG parameters in Eq. \eqref{TBGmodel},
\begin{eqnarray}
\label{mapping1}
\Omega_D&\approx& \frac{a^2|\bm{K}|^2}{\sqrt{1+3\gamma}}\left(1+\frac{\kappa}{2|a\bm{K}|^4}\right)\\
\label{mapping2}
v_0&\approx&\frac{t_0}{\hbar}\frac{a^2|\bm{K}|}{\sqrt{1+3\gamma}}\left(1-\frac{\kappa}{2|a\bm{K}|^4}\right)\\
\label{mapping3}
t_\perp &\approx&t_0\frac{\kappa}{2|a\bm{K}|^2\sqrt{1+3\gamma}}
\end{eqnarray}
This mapping can be used to obtain a TBG model equivalent to a given double-plate model. The green lines in Fig. \ref{fig:2} show the precision of the mapping at large angles. Deviations between the two are attributed to $\mathcal{O}(\kappa^2)$ and particle-hole asymmetry corrections.

\begin{figure*}
   \centering
   \includegraphics[width=0.85\linewidth]{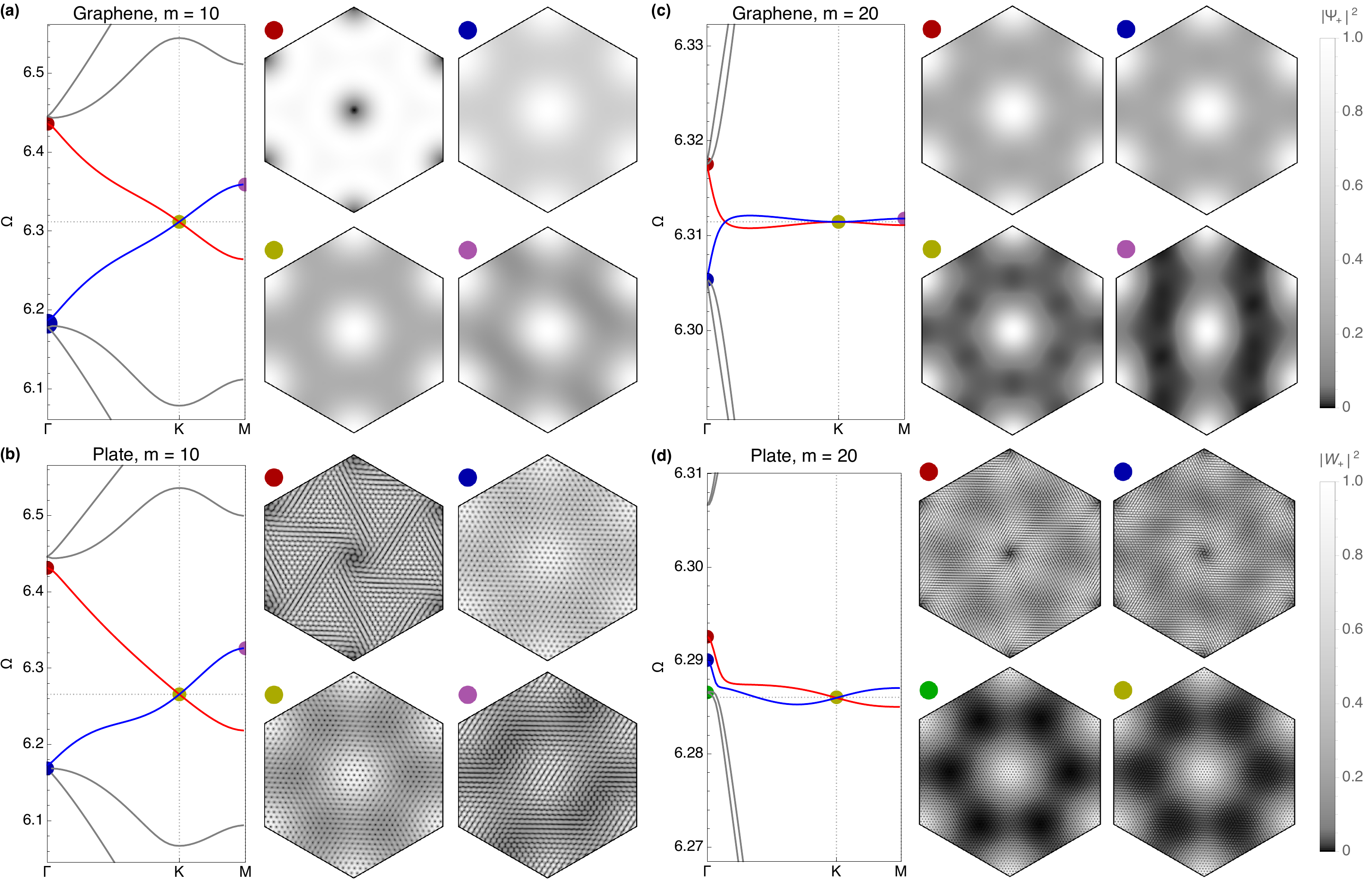} 
   \caption{Low energy (frequency) bands and spatial profile of selected eigenmodes (colored dots) for twisted bilayer graphene (a,c) and the analogous vibrating plate system (b,d). Twist angles are taken at $m=10$ (a,b) and at the first magic angle $m=20$ (c,d). Dimensionless parameters are $\kappa=40$ and $\gamma=2.5$.}
   \label{fig:4}
\end{figure*}

We now demonstrate that, as suggested by the above mapping, the structured double plates indeed develops flat bands as the angle $\theta$ is reduced, just like TBG. Figure \ref{fig:3}a shows the moir\'e superlattice and normalized bandstructure of the double plate and corresponding TBG systems, as the angle $\theta$ decreases from $\theta=7.34^\circ$ (commensurate index $m=4$) to $\theta=1.61^\circ$ ($m=20$). The latter corresponds approximately to the first magic angle $\theta_1$ for plates with $\kappa=30, \gamma=2.5$, which corresponds to a TBG with $t_\perp=0.79$eV (quite larger than in real TBG, which have $t_\perp\approx 0.48$eV, and a first magic angle at $m\approx 31, \theta\approx 1.05^\circ$.) As $\theta$ approaches $\theta_1$, the $M$-point band anticrossing grows, flattens the Dirac cones and reduces the Dirac-point group velocity $v(\theta)$. At precisely $\theta=\theta_1$ (rightmost panel in Fig. \ref{fig:3}a) the Dirac cones collapse into a quasi-flat band with exactly zero group velocity. Reducing the angle further leads to repeated reemergences and collapses of the Dirac cones at subsequent, higher-order magic angles $\theta_i$. The corresponding non-monotonous $v(\theta)$ is shown in Fig. \ref{fig:3}b, both for the double plate (blue) and TBG (green). Again, both results are very similar, with deviations again attributed to higher-order coupling corrections and particle-hole asymmetries.

We finally compare the spatial profile of the corresponding plate modes to their electron eigenstate counterparts. In magic-angle TGB, the wavefunction of flat-band electrons is predicted to be algebraically localized around AA for all momenta away from the $\Gamma$ point \cite{Laissardiere:NL10}. The fact that the states are not exponentially localized is a remarkable feature that highlights the non-trivial chirality-driven nature of the flat-band mechanism in this system. In Fig. \ref{fig:4} we compare the spatial profile of eigenmodes in twisted double plates and TBG.

At twist angles above the first magic angle, the graphene eigenstates exhibit the first hints of AA-region localization. Fig. \ref{fig:4}a shows a blowup of the bands at $\theta=3.15^\circ$ ($m=10$) in TBG. On their right, we show a selection of eigenstate top-layer densities summed over sublatices for the wavectors marked with colored circles.  States at the $K$ and $M$ points (yellow and purple dots, respectively) exhibit maxima at the AA regions (center of the hexagonal unit cell). At the $\Gamma$ point, however, the states have a different character. If we select a single $\Gamma$ state belonging to the red subband (red dot), it exhibits a minimum at AA. However, in the simplest version of the continuum model used here, $\Gamma$-point states are triply degenerate. If we plot the total spatial density from the three states, (blue dot), the AA minimum is washed out. 

The corresponding behavior of $W_+(\r)$ in the double plate system, Fig. \ref{fig:4}b is very similar. The main difference is a lack of energy symmetry around the Dirac point and, notably, the much richer spatial structure present in the eigenmodes, which, unlike for graphene, is well defined throughout the honeycomb unit cell (not only at tight-binding atomic sites). This results in an intricate fast modulation in the moir\'e supercell that reveals the chiral character of some states (e.g. red dot). The slow envelope of the different eigenmodes, however, quantitatively replicates their graphene counterparts.

At the first magic angle in graphene ($m=20$ here), the AA-localization within the flat part of the band becomes fully developed, see yellow and purple states in Fig. \ref{fig:4}c. $\Gamma$ states (red and blue) are still far less localized, and remain triply-degenerate. This is markedly different in the plate system. A true gap opens asymmetrically around the flat band, which no longer connects with higher-frequency bands at $\Gamma$. All states away from $\Gamma$ are again algebraically localized at AA (blue dot). At $\Gamma$ there are several non-degenerate states close in frequency, exhibiting varying forms of spatial structure, including AA localization (green dot). 
 
\sect{Conclusion} Our results demonstrate that the chirality-driven flat band formation mechanism of TBG can be realized in \edit{a classical system of patterned vibrating plates, see Fig. \ref{fig:1}. We derived a quantitatively precise mapping between the graphene-based and mechanical systems, and demonstrated a very similar modulation of group velocity with twist angle and spatial eigenmode profiles in both}. The differences in spectral properties are mostly due to (a) the increased number of spatial degrees of freedom in the plate as compared to the tight-binding graphene models, and (b) the simplified form of the graphene model, which here does not incorporate perturbations that break particle-hole symmetry and $\Gamma$-point degeneracies. The exploration of the magic angle sequence and velocity modulation of TBG using the mechanical double-plate analogue would allow far easier parameter uniformity and control than in TBG. Mechanical analogues could not only help shed light on the rich TBG physics, but would also enable ultrasonics devices for slow-sound operations and RF signal buffering. 

\acknowledgements

F.P and P.S-J. acknowledge support from the Spanish Ministry of Science, Innovation and Universities through Grants PCI2018-093026 and PGC2018-097018-B-I00 (AEI/FEDER, EU). J.C. acknowledges the support from the European Research Council (ERC) through the Starting Grant 714577 PHONOMETA and from the MINECO through a Ram\'on y Cajal grant (Grant No. RYC-2015-17156).

\appendix

\section{Equation of motion of the double plate system}
\label{ap:1}

The real-space equation of motion of the double plate system reads, in the frequency domain
\begin{eqnarray}
\sum_{l'}\left[(h\rho\omega^2-D\nabla^4 - \mathcal{K})\tau_0^{ll'} + \mathcal{K}\tau_x^{ll'}\right]W_{l'}(\r,\omega) \nonumber\\
= -h\omega^2\delta\rho_l(\r) W_l(\r,\omega)
\end{eqnarray}
with 
\[
\delta\rho_l(\r)=\sum_{\alpha}\sum_{\r^l_\alpha} \frac{\Delta m}{A_c h}\delta(\r-\r^l_{\alpha})
\]
Layer index is $l=1,2$, with $\tau_i$ the corresponding Pauli matrices, sublattice index is $\alpha=A,B$, flexural field is $W_l$, $A_c$ is the honeycomb unitcell area, $\mathcal{K}$ is the interlayer elastic coupling constant, $\Delta m$ is the mass defect, and defect positions $\r^l_{\alpha}$ are
\begin{eqnarray}
\r^l_{\alpha} &=& \r^l_0 + \dr^l_\alpha\nonumber\\
\r^l_0 &=& n_1 \bm{a}^l_1+n_2 \bm{a}^l_2\nonumber\\
\dr^l_{A,B} &=& \mp (\bm{a}^l_1+\bm{a}^l_2)/6
\label{drAB}
\end{eqnarray}
Here $\r^l_0$ is the center of a honeycomb unit cell, of period $a=|\bm{a}_i^l|$, and $\dr^l_\alpha$ are sublattice shifts. The conjugate $\g^l_{1,2}$ of the layer-dependent Bravais vectors $\bm{a}^l_{1,2}$ are defined through $\bm{a}^l_i\cdot\g^l_j=2\pi\delta_{ij}$. 

In analogy to $\r^l_0$ we denote all integer multiples of $\g^l_i$ by
\[
\g^l = n_1 \g^l_1+n_2\g^l_2
\]
The moir\'e conjugate vectors are denoted by $\bm{G}_{1,2}$. Its integer multiples are the collection of all possible $\g^l-\g^{l'}$
\[
\bm{G} = \g^l-\g^{l'} = n_1 \bm{G}_1+n_2\bm{G}_2
\]

\subsection{Equation of motion in eigenvalue form}

The equation of motion can be written in the operator-ket language as

\begin{equation}
\label{eig}
\hat\G^{-1}\ket{w} = (\hat\G_0^{-1} - \hat\Sigma)\ket{w} = 0
\end{equation}
where
\begin{eqnarray*}
\braket{\r l}{w} &=& \frac{D}{a^4}W_l(\r,\omega)\\
\hat\G^{-1} &=& \hat\G_0^{-1} - \hat\Sigma\\
\hat\G_0^{-1} &=& \left[(\Omega^2-a^4\hat\k^4-\kappa)\tau_0 + \kappa\tau_x\right]\\
\hat\G_0 &=&\sum_{ll'}\intk \ket{\k l'}\left(\frac{1}{(\Omega^2-|a\k|^4-\kappa)\tau_0+\kappa\tau_x}\right)_{l'l}\bra{\k l}\\
\hat \Sigma &=& -\Omega^2\gamma\sum_{l\alpha\r_\alpha^l}\ket{\r_\alpha^l l}\bra{\r_\alpha^l l}
\end{eqnarray*}
Dimensionless constants are
\begin{equation*}
\Omega^2 = \frac{a^4h\rho\omega^2}{D} \hspace{0.5cm}
\kappa = \frac{a^4\mathcal{K}}{D} \hspace{0.5cm}
\gamma = \frac{\Delta m}{\rho h A_c}
\end{equation*}
and $\ket{\r l}$ and $\ket{\k l}$ are layer-resolved eigenstates of position $\hat \r$ and momentum $\hat \k$ operators, respectively, with $\braket{\r l'}{\k l}= \delta_{ll'}e^{i\k\r}$, $\braket{\k' l'}{\k l}= \delta_{ll'}(2\pi)^2\delta(\k'-\k)$ and $\braket{\r' l'}{\r l}\delta_{ll'}\delta(\r'-\r)$.

The mass defects, encoded in $\hat\Sigma$, induce momentum scattering by $\g^l$ on each layer. Using \[\sum_{\r_0^l}e^{i\k \r_0^l} = (2\pi)^2\sum_{\g^l} \delta(\k-\g^l)\] we can prove
\begin{equation}
\hat \Sigma = -\Omega^2\gamma\sum_{l\alpha\g^l}\intk e^{i\g^l\dr_\alpha^l}\ket{\k l}\bra{\k+\g^l,l}
\end{equation}
Projecting Eq. \eqref{eig} onto $\bra{\k+\bm{G},l}$ we obtain the eigenvalue equations
\begin{eqnarray}
\label{eom}
\sum_{l'}\left(\Omega^2-a^4|\k+\bm{G}|^4-\kappa)\tau^{ll'}_0 + \kappa\tau^{ll'}_x\right)\braket{\k+\bm{G},l'}{w}\\
=-\Omega^2\gamma\sum_{\bm{G}'}\braket{\k+\bm{G}',l}{w}\sum_{\alpha,\g^l}e^{i\g^l\dr^l_\alpha}\delta_{\bm{G}',\bm{G}+\g^l}\nonumber
\end{eqnarray}
We can arrange the above as a matrix eigenvalue equation
\begin{eqnarray}
\left[\Omega^2(1+\gamma \mathcal{T})-\mathcal{P}\right]w &=& 0\nonumber\\
\mathcal{P}^{-1}(1+\gamma \mathcal{T})w &=& \frac{1}{\Omega^2}w
\label{eigs}
\end{eqnarray}
where
\begin{eqnarray}
w_{l\bm{G}} &=& \braket{\k+\bm{G},l'}{w}\\
\mathcal{T}_{l'\bm{G}',l\bm{G}} &=& \tau_0^{l'l}\sum_{\g^l}2\cos(\g^l\dr^l_A)\delta_{\bm{G}',\bm{G}+\g^l}\\
\mathcal{P}_{l'\bm{G}',l\bm{G}} &=& \delta_{\bm{G}',\bm{G}}\left[(a^4|\k+\bm{G}|^4+\kappa)\tau^{ll'}_0 - \kappa\tau^{ll'}_x\right]
\end{eqnarray}
Equation \eqref{eigs} can be readily be solved using standard exact diagonalization numerical routines, which yields dispersions $\Omega(\k)$ and eigenstates $w$.

\section{Mapping between the double-plate system and a twisted bilayer graphene}

\label{ap:2}

In this section we derive analytically the connection between the double-plate equation of motion and the standard continuum model of twisted bilayer graphene. The essence of the procedure is to perturbatively integrate out the flexural plate vibrations between scatterers, and thus obtain an effective eigenvalue equation for vibrations localized at scattering sites, valid for weak interplate coupling.

\subsection{Description in the tight-binding subspace}

An alternative solution to the diagonalization of Eq. \eqref{eom} relies on using Dyson's equation 
\begin{equation}
\label{Dyson}
\hat\G = \hat \G_0 + \hat\G_0\hat\Sigma\hat\G
\end{equation}
to obtain $\hat \G$, and then finding the nullspace of its inverse, Eq. \eqref{eig}. We can formally solve  Eq. \eqref{Dyson} using the $\T$-matrix approach
\begin{eqnarray*}
\hat\G &=& \hat\G_0+\hat\G_0 \hat\T \hat\G_0\\
\hat\T &=& \hat\Sigma + \hat\Sigma \hat\G \hat\Sigma
\end{eqnarray*}
The nullspace of $\hat\G^{-1}$ can be obtained from the nullspace of $\hat \T^{-1}$, or alternatively the nullspace of $(\hat\Sigma \hat\G \hat\Sigma)^{-1}$. The key advantage is that since $\hat\Sigma$ projects onto the subspace of discrete positions of the honeycomb lattices, we only need to invert the projection $\hat\G^P$ of $\hat\G$ on said subspace, 
\begin{eqnarray}
\hat\G^P &=& \hat\G_0^P + \hat\G_0^P\hat\Sigma^P\hat\G^P\\
(\hat\G^P)^{-1}  &=& (\hat\G_0^P)^{-1} -\hat\Sigma^P
\label{GP}
\end{eqnarray}
where we denote
\begin{eqnarray*}
\hat\G^P &=& \hat P\hat\G\hat P\\
\hat\G_0^P &=& \hat P\hat\G_0\hat P\\
\hat\Sigma^P &=& \hat\Sigma =  -\Omega^2\gamma \hat P\\
\end{eqnarray*}
and where $\hat P$ denotes the tight-binding projector,
\begin{eqnarray*}
\hat P &=& \sum_{l\alpha}\hat P_\alpha^l\\
\hat P_\alpha^l &=& \sum_{\r_\alpha^l}\ket{\r_\alpha^l}\bra{\r_\alpha^l}\\&=&\int_{\textrm{BZ}_l} \frac{d^2k_0}{(2\pi)^2}\ket{\k_0l\alpha}\bra{\k_0l\alpha}
\end{eqnarray*}
In the last equality we have introduced tight-binding plane waves, i.e. plane waves projected on a given honeycomb sublattice and layer
\[
\ket{\k_0l\alpha} = e^{-i\k_0\dr_\alpha^l}\hat P_\alpha^l\ket{\k_0 l} = \sum_{\r_0^l}e^{i\k_0\r_0^l}\ket{\r_0^l+\dr_\alpha^l}
\]
Note that $\ket{\k_0l\alpha} = \ket{\k_0+\g^l,l\alpha}$, i.e. $\k_0$ is the tight-binding wavevector, only defined modulo $\g^l$, within the honeycomb's first Brillouin zone of layer $l$ (denoted BZ${}_l$). 

Tight-binding plane waves project onto continuum plane waves as
\begin{equation}
\langle \bm{k}l|\bm{k}_0\alpha l\rangle = \sum_{\bm{g}^l}(2\pi)^2\delta(\bm{k}-\bm{k}_0-\bm{g}^l)e^{-i\bm{k}\delta\bm{r}^l_\alpha}
\end{equation}

\begin{widetext}
This allows us to arrive at a crucial result that the projected $\hat\G^P_0$ only mixes $\k_0$'s that differ by a moir\'e wavevector $\bm{G}=\g^{l'}-\g^l$

\begin{eqnarray}
\G_0^{\k_0'l'\alpha',\k_0l\alpha}&\equiv& \bra{\k_0'l'\alpha'}\hat\G^P_0\ket{\k_0l\alpha}=\bra{\k_0'l'\alpha'}\hat\G_0\ket{\k_0l\alpha}\nonumber\\
&=& \sum_{\bm{g}^{l'}\bm{g}^l}(2\pi)^2\delta\left((\k_0'-\k_0)-(\bm{g}^{l}-\bm{g}^{l'})\right) 
e^{i(\k_0+\g^l)(\dr^{l'}_{\alpha'}-\dr^l_{\alpha})}
\left(\frac{1}{(\Omega^2-a^4|\k_0+\g^l|^4-\kappa)\tau_0+\kappa\tau_x}\right)_{l'l}
\label{GTB}
\end{eqnarray}

For a conmensurate rotation angle $\theta = \arccos\left[(3m^2 + 3m + 1/2)/(3m^2 + 3m + 1)\right]$ (arbitrary integer index $m$) between the top layer $l'=+$ and the bottom layer $l=-$, we have
\begin{eqnarray}
\bar{\bar{G}}\equiv\left(\begin{array}{c}\bm{G}_1 \\ \bm{G}_2\end{array}\right) = 
M^{-1}
 \left(\begin{array}{c}\g^-_1 \\ \g^-_2\end{array}\right) = 
(M^{-1})
^T 
\left(\begin{array}{c}\g^+_1 \\ \g^+_2\end{array}\right),
\hspace{.5cm}
M=\left(\begin{array}{cc}2m+1 & -m -1\\ -m & 2m+1\end{array}\right)\end{eqnarray}

It is possible to show from the above that for opposite layers $l=-, l'=+$, the $\g^{+},\g^-$ that correspond to any given $\bm{G}= \nu_1 \bm{G}_1+\nu_2 \bm{G}_2=\g^{-}-\g^+$ are unique and equal to $\g^- = -\nu_2 \g^-_1 + \nu_1 \g^-_2$, $\g^{+} = -\nu_2 \g^{+}_1 + \nu_1 \g^{+}_2$. For intralayer $l=l'$ elements, $\bm{G}=0$, since $\g^l=\g^{l'}$ (recall that $\k_0$ lies in the honeycomb's first Brillouin zone). We can use this result in the $l\neq l'$ case to write $\g^- = \bm{\nu}N\bar{\bar{G}}$ for a given $\bm{G}=\g^{-}-\g^+=\bm{\nu}\bar{\bar{G}}$, where $N = \left(\begin{array}{cc}0 & 1\\ -1 & 0\end{array}\right)M$

Decomposing $\hat\G_0^P$ into intralayer $\hat\G_0^\parallel$ and interlayer $\hat\G_0^\perp$ components, we finally get
\begin{eqnarray}
\label{GTB2}
\hat\G_0^P &=& \hat\G_0^\parallel+\hat\G_0^\perp\\
\hat\G_0^\parallel &=& \sum_{\alpha\alpha'l=\pm}\sum_{\g^l}\int_{\textrm{BZ}_l}\frac{d^2k_0}{(2\pi)^2}
\ket{\k_0\alpha' l} 
e^{i(\k_0+\g^l)(\dr^{l}_{\alpha'}-\dr^l_{\alpha})}
\left(\frac{\Omega^2-a^4|\k_0+\g^l|^4-\kappa}{(\Omega^2-a^4|\k_0+\g^l|^4-\kappa)^2-\kappa^2}\right)\bra{\k_0\alpha l}\nonumber\\
\hat\G_0^\perp &=& \sum_{\alpha\alpha'}\sum_{\bm{\nu}}\int_{\textrm{BZ}_-}\frac{d^2k_0}{(2\pi)^2}
\ket{\k_0+\bm{\nu}\bar{\bar{G}},\alpha'+} 
e^{i(\k_0+\bm{\nu}N\bar{\bar{G}})(\dr^{+}_{\alpha'}-\dr^-_{\alpha})}
\left(\frac{-\kappa}{(\Omega^2-a^4|\k_0+\bm{\nu}N\bar{\bar{G}}|^4-\kappa)^2-\kappa^2}\right)\bra{\k_0\alpha -} + \mathrm{h.c.}\nonumber
\end{eqnarray}
where $\bar{l}$ stands for the layer opposite to $l$.

\end{widetext}

\subsection{Decoupled layers}
An important limit to consider for Eq. \eqref{GTB2} is that of decoupled layers, $\kappa=0$. In this limit, $\hat\G_0^\perp=0$, and $\k_0$ is preserved. We are left with a $2\times 2$ matrix in sublattice space
\begin{eqnarray}
\G_0^{\k_0'l'\alpha',\k_0l\alpha} &=& (2\pi)^2\delta\left(\k_0'-\k_0\right) \delta_{ll'}\,\G_0^{\alpha'\alpha}(\k_0,l)\nonumber\\
\G_0^{\alpha'\alpha}(\k_0, l)&=&\sum_{\g^l}\frac{e^{i(\k_0+\g^l)(\dr_{\alpha'}-\dr_{\alpha})}}{\Omega^2-a^4|\k_0+\g^l|^4}
\label{G2x2}
\end{eqnarray}
In this case Eq. \eqref{GP} factors into exactly solvable $2\times 2$ blocks, 
\begin{equation}
\left[\G^P(\k_0, l)\right]^{-1}  = \left[\G_0^P(\k_0, l)\right]^{-1} -\Sigma^P(\k_0, l)
\label{DysonP}
\end{equation}
Here the $\Sigma^P(\k_0, l)$ matrix is diagonal, $\Sigma^{\alpha'\alpha}(\k_0, l) = -\Omega^2\gamma\delta_{\alpha'\alpha}$ and the $\G_0^P(\k_0, l)$ matrix is given by Eq. \eqref{G2x2}. Expanding the above to linear order around specific values $\k_0=\bm{K}^l=(\g^l_1-\g^l_2)/3$ and $\Omega_D$ (Dirac point) leads to a low energy Dirac Hamiltonian of the form
\begin{eqnarray}
\left[\G^P(\k_0, l)\right]^{-1} &\approx& \beta\left[(\Omega-\Omega_D)\sigma_0 - v\,(\k_0-\bm{K}^l)\cdot\bm{\sigma}\right]\\
&=&\beta\left[(\Omega-\Omega_D)\sigma_0 - H_\textrm{Dirac}(\k_0-\bm{K}^l)\right]\nonumber
\label{Dirac} 
\end{eqnarray}
where $\bm{\sigma}=(\sigma_x, \sigma_y)$ are Pauli matrices in sublattice space. This form is obtained by equating Eq. \eqref{DysonP} at $\Omega=\Omega_D$ and $\k_0=\bm{K}^l$ to zero, i.e. $\G^P_0(\bm{K}_l, l)\Sigma^P(\bm{K}_l, l)=\sigma_0$.
The off-diagonal elements are exactly zero at $\k_0=\bm{K}^l$. The diagonal elements become zero at a given $\Omega_D$, that thus satisfies the non-linear equation
\begin{equation}
\G_0^{\alpha\alpha}(\bm{K}_l, l) = -\frac{1}{\Omega_D^2\gamma} 
\label{OmegaD}
\end{equation}
or more explicitly
\begin{equation}
\sum_{\g^l}\frac{1}{\Omega_D^2-a^4|\bm{K}^l+\g^l|^4} = -\frac{\rho h A_c}{\Omega_D^2\Delta m}
\label{OmegaD2}
\end{equation}
Its solution $\Omega_D$ is the same in both layers $l$.

The value of the constant $\beta$ is obtained from
\begin{eqnarray}
\beta &=& \frac{1}{2}\mathrm{Tr}\left[\sigma_0\partial_{\Omega}\left[\G^P(\k_0,l)\right]^{-1}\right]_{\k_0=\bm{K}, \Omega=\Omega_D} 
\end{eqnarray}

The Dirac velocity $v_0=v_x=v_y$ (isotropic) is obtained by differentiating Eq. \eqref{Dirac},
\begin{eqnarray}
v_i&=&-\frac{1}{2\beta}\mathrm{Tr}\left[\sigma_i \partial_{\k_0^i}\left[\G^P(\k_0,l)\right]^{-1}\right]_{\k_0=\bm{K}, \Omega=\Omega_D} \nonumber\\
&=& \frac{\Omega_D^4\gamma^2}{2\beta}\mathrm{Tr}\left[\sigma_i \partial_{\k_0^i}\G^P_0(\k_0,l)\right]_{\k_0=\bm{K}, \Omega=\Omega_D}
\label{vD}
\end{eqnarray}
The latter is a consequence of Eq. \eqref{OmegaD2}. Then
\begin{eqnarray}
\label{vD2}
|v_i|&=& \frac{4a^4\Omega_D^4\gamma^2}{\beta} \left|\sum_{\g^l}\frac{e^{i(\bm{K}^l+\g^l)(\bm{a}^l_1+\bm{a}^l_2)/3}|\bm{K}^l+\g^l|^2}{(\Omega_D^2-a^4|\bm{K}^l+\g^l|^4)^2}(\bm{K}^l+\g^l)_i\right|\nonumber
\end{eqnarray}

A good approximation is to truncate the $\sum_{\g^l}$ to the three larger terms, which are actually equal. These are $\g^l=\{0,-\g^l_2,\g^l_1\}$. For all these, $|\bm{K}^l+\g^l|$ is equal and has minimal value $|\bm{K}|=4\pi/(3a)$. This is the `first star' approximation. It yields
\begin{eqnarray}
\label{OFS}
\Omega_D&\approx&\frac{a^2|\bm{K}|^2}{\sqrt{1+3\gamma}}\\
v_0&\approx&\frac{a^2|\bm{K}|}{\sqrt{1+3\gamma}}\\
\beta &\approx& \frac{2}{3}\sqrt{1+3\gamma}|a\bm{K}|^2
\label{vFS}
\end{eqnarray}
Recall that $\gamma=\Delta m/(\rho h A_c)$. Note that $\gamma>-1/2$, since although $\Delta m$ can be negative (hole defects), each defect cannot remove more mass than $m_c=\rho h A_c$, the mass the plate has in half a unit cell. However, a critical window \[-1/2<\gamma<-1/3,\] seems to exist that is physically possible and for which there is no Dirac point solution.

\subsection{Coupled layers}

For coupled layers ($\kappa\neq 0$) both $\hat\G_0^\perp$ and $\hat\G_0^\parallel$ are non-zero. Moreover, since $\hat\G_0^\perp$ couples $\ket{\k_0l}$ to $\ket{\k_0+\bm{G},\bar{l}}$ we can no longer decompose the problem into a $2\times 2$ matrix form for a fixed $\k_0$. 

Let us consider $\hat\G_0^\parallel$ first. It contributes to the low-energy Hamiltonian much like in Eq. \eqref{Dirac}, with a Dirac Hamiltonian around $\Omega=\Omega_D$ and $\k_0=\bm{K}^l$ on each layer $l$. The finite $\kappa$ however changes the value of $\Omega_D$ and $v$, that still satisfy Eqs. \eqref{OmegaD} and \eqref{vD}, with $\G^P_0(\k_0,l)$ replaced by $\G^\parallel_0(\bm{\k_0,l}) = \bra{\k_0l}\hat\G^\parallel_0\ket{\k_0l}$, see Eq. \eqref{GTB2}. First-star approximations yield more complicated though still analytical solutions. In particular we have

\begin{eqnarray}
\Omega_D&\approx& \frac{|a\bm{K}|^2}{\sqrt{1+3\gamma}}\left(1+\frac{\kappa}{2|a\bm{K}|^4}\right)+\mathcal{O}(\kappa^2)\\
\label{Okappa}
v_0&\approx&\frac{a^2|\bm{K}|}{\sqrt{1+3\gamma}}\left(1-\frac{\kappa}{2|a\bm{K}|^4}\right) + \mathcal{O}(\kappa^2)\nonumber\\
\beta &\approx& \frac{2}{3}\sqrt{1+3\gamma}|a\bm{K}|^2\left(1+\frac{\kappa}{2|a\bm{K}|^4}\right) + \mathcal{O}(\kappa^2)\nonumber
\end{eqnarray}

To obtain the interlayer contribution to the low energy Hamiltonian we would need to compute
\begin{equation}
\left[\hat\G^P\right]^{-1}  = \left[\hat\G_0^\parallel+\hat\G_0^\perp\right]^{-1} -\hat\Sigma^P
\end{equation}
This can no longer be cast into a matrix equation of finite-size. Instead we can now think of the matrices to be discrete but infinite, in the subspace spanned by $\ket{\k_0+\bm{G},\alpha,l}$ for all $\bm{G},\alpha,l$. Inverting $\hat\G_0$ is now no longer possible to do analytically. We can however still do it approximately using perturbation theory in the interlayer coupling $\kappa$,
\begin{eqnarray}
\left[\hat\G^P\right]^{-1}  &=& \hat\G_0^{\parallel^{-1}} -\hat\Sigma^P \\
&&-\hat\G_0^{\parallel^{-1}}\hat\G_0^\perp\,\hat\G_0^{\parallel^{-1}}
+\hat\G_0^{\parallel^{-1}}\hat\G_0^\perp\,\hat\G_0^{\parallel^{-1}}\hat\G_0^\perp\,\hat\G_0^{\parallel^{-1}} -...\nonumber
\end{eqnarray}
Note that the inverse $\hat\G_0^{\parallel^{-1}}$ is still easy because it does not mix $\k_0$. Note also that the even-order-in-$\hat\G_0^\perp$ terms renormalize the intralayer Hamiltonian with $\k_0$-scattering contributions, while the odd-order terms contribute to the interlayer coupling. To arrive at a Hamiltonian similar to that of twisted-layer graphene we must therefore stop at linear order, and neglect $\mathcal{O}(\kappa^2)$ contributions. This approximation is valid for $\kappa$ much smaller than the associated scale in the Hamiltonian, i.e. $|a\bm{K}|^4$, so for $\kappa\ll 308$. In this weak-coupling limit we get
\begin{widetext}
\begin{equation}
H_\textrm{eff}(\k_0) = \left(\begin{array}{cc}
H_\textrm{Dirac}(\k_0-\bm{K}^-+\bm{G})\delta_{\bm{G}\bm{G'}} &
\left[H^\perp_{\bm{G}'\bm{G}}(\k_0)\right]^\dagger\\
H^\perp_{\bm{G}'\bm{G}}(\k_0)
& H_\textrm{Dirac}(\k_0-\bm{K}^++\bm{G})\delta_{\bm{G}\bm{G'}}
\end{array}\right)
\end{equation}
where blocks correspond to $l=-$ (bottom) and $l=+$ (top) layers. The Dirac Hamiltonians contain the $\Omega_D$ and $v$ parameters in the  $\mathcal{O}(\kappa)$ `first star' approximation, Eqs. \eqref{Okappa}. The interlayer coupling reads
\[
\left[H^\perp_{\bm{G}'\bm{G}}(\k_0)\right]^{\alpha'\alpha}= \frac{1}{\beta}\left.\bra{\k_0+\bm{G}',\alpha',+}\hat\G_0^{\parallel^{-1}}\hat\G_0^\perp\,\hat\G_0^{\parallel^{-1}}\ket{\k_0+\bm{G},\alpha,-}\right|_{\Omega=\Omega_D}
\]
Since $\bra{\k_0,\alpha',l}\hat\G_0^{\parallel^{-1}}\ket{\k_0,\alpha,l}\approx\bra{\k_0,\alpha',l}\hat\Sigma^P\ket{\k_0,\alpha,l}=-\Omega_D^2\gamma\delta_{\alpha\alpha'}$ at $\Omega=\Omega_D$ and close to $\k_0=\bm{K}^l$, we have
\begin{eqnarray}
\left[H^\perp_{\bm{G}'\bm{G}}(\k_0)\right]^{\alpha'\alpha} &=& \frac{\Omega_D^4\gamma^2\kappa}{\beta}\frac{e^{i(\k_0+\bm{\nu}N\bar{\bar{G}})(\dr^+_{\alpha'}-\dr^-_{\alpha})}}{(\Omega_D^2-a^4|\k_0+\bm{\nu}N\bar{\bar{G}}|^4)^2},\hspace{0.5cm}
N=\left(\begin{array}{cc}
-m & 2m+1\\-2m-1 & m+1
\end{array}\right)
\end{eqnarray}
where $\bm{G}' - \bm{G} = \bm{\nu}\bar{\bar{G}}$.
\end{widetext}

A small-angle approximation can be made for large $m$. In this case we can drop the dependence on $l$ of $\dr^l_\alpha$, $\g^l$ and $\bm{K}^l$, and make $\k_0=\bm{K}$. If we furthermore adopt the first star approximation we can drop all $\bm{\nu}$ except $\bm{\nu} =\{(0,0),(1,0),(0,1)\}$. These correspond to $\bm{G}'-\bm{G}= \bm{\nu}\bar{\bar{G}} = \{0,\bm{G}_1, \bm{G}_2\}$ and $\g = \bm{\nu}N\bar{\bar{G}} = \{0, \g_2, -\g_1\}$.
Under these approximations,
\begin{eqnarray}
\left[H^\perp_{\bm{G}'\bm{G}}(\k_0)\right]^{\alpha'\alpha} &\approx&\frac{\Omega_D^4\gamma^2\kappa}{\beta(\Omega_D^2-|a\bm{K}|^4)^2}\tau^{\bm{\nu}}_{\alpha'\alpha}\nonumber\\
&=&\frac{\kappa}{6|a\bm{K}|^2\sqrt{1+3\gamma}}\tau^{\bm{\nu}}_{\alpha'\alpha}\nonumber
\end{eqnarray}
Using Eq. \eqref{drAB}, matrix $\tau^{\bm{\nu}}_{\alpha'\alpha}$ reads
\begin{eqnarray}
\tau^{\bm{\nu}}_{\alpha'\alpha} &=& e^{i(\bm{K}+\bm{\nu}N\bar{\bar{G}})(\dr_{\alpha'}-\dr_{\alpha})} \\
\tau^{(0,0)} &=& \left(\begin{array}{cc}
1 & 1\\1 & 1
\end{array}\right)\nonumber\\
\tau^{(1,0)} &=& \left(\begin{array}{cc}
1 & e^{-i2\pi/3}\\e^{i2\pi/3} & 1
\end{array}\right)\nonumber\\
\tau^{(0,1)} &=& \left(\begin{array}{cc}
1 & e^{i2\pi/3}\\e^{-i2\pi/3} & 1
\end{array}\right)\nonumber
\end{eqnarray}
The above reproduces the interlayer coupling of the continuum model for twisted bilayer graphene, Eq. \eqref{TBGmodel}, with
\begin{eqnarray}
v_0&\approx&\frac{t_0}{\hbar}\frac{a^2|\bm{K}|}{\sqrt{1+3\gamma}}\left(1-\frac{\kappa}{2|a\bm{K}|^4}\right)\\
t_\perp &=& \frac{t_0\kappa}{6|a\bm{K}|^2\sqrt{1+3\gamma}}
\end{eqnarray}
where we have restored the energy scale $t_0\approx 2.7$eV characteristic of graphene's.

\bibliography{biblio}

\end{document}